\documentclass[preprint,preprintnumbers,aps,amsmath,amssymb,prl,superscriptaddress,floatfix]{revtex4-1}

\usepackage{graphicx}
\usepackage{dcolumn}
\usepackage{bm}
\usepackage[T1]{fontenc}
\usepackage[utf8]{inputenc}

\begin{document}

\title{Tunneling into the vortex state of NbSe$_2$ with van der Waals junctions}

\author{T. Dvir}
\affiliation{The Racah Institute of Physics, the Hebrew University of Jerusalem, Israel}

\author{M. Aprili}
\affiliation{Laboratoire de Physique des Solides (CNRS UMR 8502), Bâtiment 510, Université Paris-Sud/Université Paris-Saclay, 91405 Orsay, France}

\author{C. H. L. Quay}
\affiliation{Laboratoire de Physique des Solides (CNRS UMR 8502), Bâtiment 510, Université Paris-Sud/Université Paris-Saclay, 91405 Orsay, France}

\author{H. Steinberg}
\affiliation{The Racah Institute of Physics, the Hebrew University of Jerusalem, Israel}

\date{\today}

\begin{abstract}
We have performed device-based tunnelling spectroscopy of NbSe$_2$ in the vortex state with a magnetic field applied both parallel and perpendicular to the $a-b$ plane. Our devices consist of layered semiconductors placed on top of exfoliated NbSe$_2$ using the van der Waals transfer technique. At zero field, the spectrum exhibits a hard gap, and the quasiparticle peak is split into low and high energy features. The two features, associated with the effective two-band nature of superconductivity in NbSe$_2$, exhibit markedly distinct responses to the application of magnetic field, suggesting an order-of-magnitude difference in the spatial extent of the vortex cores of the two bands.
At energies below the superconducting gap, the hard gap gives way to vortex-bound Caroli-de Gennes-Matricon states, allowing the detection of individual vortices as they enter and exit the junction. Analysis of the sub-gap spectra upon application of parallel magnetic field allows us to track the process of vortex surface formation and spatial rearrangement in the bulk. 
\end{abstract}


\maketitle

\subsection*{Introduction}

Magnetic fields can penetrate Type-II superconductors through normal regions (vortex cores) around which supercurrents circulate; this mixed or vortex state occurs between the superconductor’s lower and upper critical magnetic fields. 
Tunnelling spectroscopy of the vortex state was first performed by Hess et al.~\cite{Hess1989a} on 2H-NbSe$_2$ (hereafter NbSe$_2$). They and others~\cite{Hess1991,Guillamon2008,Suderow2014a} observed bound states in vortex cores predicted by Caroli, de Gennes, and Matricon~\cite{Caroli:1964fe}, which form a quasi-continuous spectrum at energies below the superconducting gap $\Delta$. Like other transition metal dichalcogenides (TMDs), NbSe$_2$ has a layered crystal structure and cleaves easily perpendicular to its \textit{c}-axis. Recently, NbSe$_2$ has seen a revival of interest due to the availability of van der Waals mechanical integration methods, allowing the fabrication of transport~\cite{Xi2015} and tunneling~\cite{Amet2012,Dvir:2018fj,Khestanova:2018fs,Costanzo:2018bv} devices incorporating flakes of varying thicknesses, down to the monolayer. 

In the NbSe$_2$ tunneling spectrum~\cite{Guillamon2008,Noat2015,Dvir:2018fj}, the quasiparticle feature peaking at $\Delta=1.25$ meV exhibits a shoulder at 600 $\mu eV$. This splitting, evident in the 2$^{nd}$ derivative of the tunneling current, is associated with an effective two-band model. The structure of vortices in a two-band superconductor is not trivial~\cite{Ichioka:2004hf}. In MgB$_2$, a well studied two-band superconductor, the two quasiparticles peaks are well separated in energy ($\Delta_{1,2}=$2.2,7 meV, respectively). The well-resolved peaks, allow tracking their distinct responses to the onset of the vortex state~\cite{Eskildsen:2003kw}: while the higher energy quasiparticle peak is recovered within a short distance from the vortex core, corresponding to a short coherence length, the lower energy feature recovers over a larger length scale, and within a finite magnetic field it disappears altogether. 

Observing similar effects in NbSe$_2$ is challenging due to the small energy separation between the two gaps. In a previous study~\cite{Dvir:2018fj}, we showed that by measuring tunnel devices at millikelvin temperatures it is possible to differentiate the two spectral features. When following their response to Meissner currents, we found that the small energy feature exhibits very fast depairing, suggesting that the two bands have very different dynamical properties. Such a difference is expected to lead to substantial changes in coherence length, and hence in the spatial extent of vortex cores.

\begin{figure}
\includegraphics[width = 0.9\textwidth]{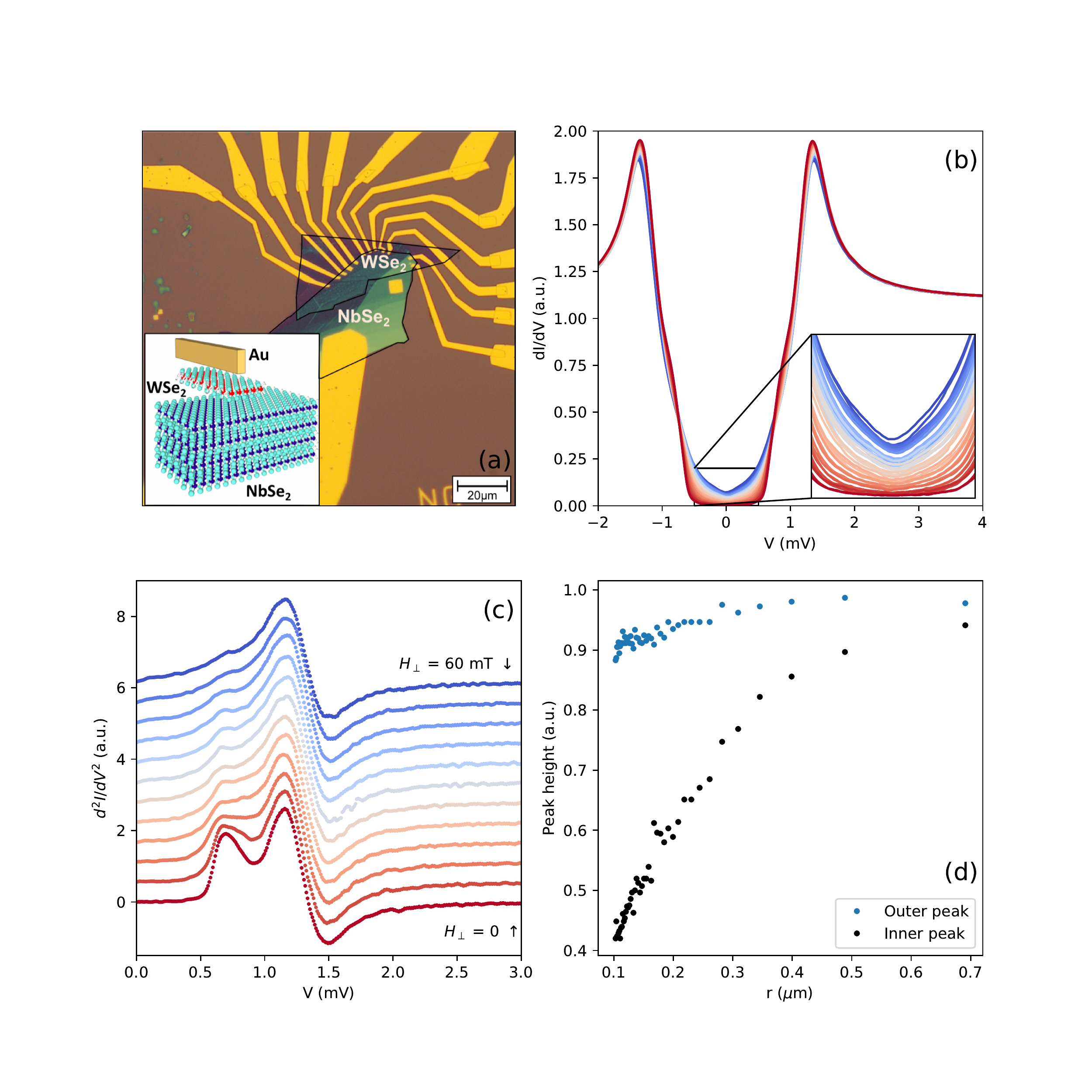}
\caption{Spectral response of NbSe$_2$ to out of plane magnetic field. (a.) An optical image of one of the devices. Thin WSe$_2$ is placed on top of a NbSe$_2$ flake of variable thickness. The inset shows a schematic of a tunnel junction. (b.) $dI/dV$ vs. $V$ of junction 2 ($d$ = 25 nm) normalized to the tunneling conductance at the normal state. Different curves are taken at increasing $H_\perp$. Inset: closeup on the evolution of the sub gap spectra. Colors range between red (zero field) and blue ($H_\perp$ = 60 mT). (c.) Same for $dI^2/dV^2$, curves shifted vertically for clarity. (d.) The height of the two peaks observed in the second derivative as a function of the distance between the vortices. } 
\label{PerpField}
\end{figure}

In this work, we perform tunnelling spectroscopy of NbSe$_2$ using Normal-Insulator-Superconductor (NIS) tunnel junctions. The junctions are fabricated by placement of MoS$_2$ or WSe$_2$ tunnel barriers on top of a NbSe$_2$ flake, using the van der Waals (vdW) transfer technique. Electrodes and leads are fabricated following the procedure reported earlier~\cite{Dvir:2018fj}. Typical junction dimensions are in the of 1-2 $\mu m^2$ (Figure \ref{PerpField} (a)) and typical barriers consisting of 3-5 layers of TMD's giving a thickness of 2-3 nm.  A bias voltage $V$ is applied across the NIS tunnel junction, and the current $I$ and differential conductance $G = dI/dV$ measured using standard lock-in techniques. At low temperatures, the G(V) of NIS junctions is proportional to the density of states (DOS) $N(eV)$ of the superconductor \cite{tinkham2004introduction}, with $e$ the electron charge. We report tunnel measurements from 7 junctions. The tunneling spectra, taken at $T=70$~mK (unless otherwise stated), show that the two components of the quasiparticle peak exhibit distinct responses to the onset of a perpendicular magnetic field. Also, we can track the spectral signature of the spatially integrated vortex-bound CdGM states, and their evolution in magnetic fields applied both in and perpendicular to the $a-b$ plane.

\subsection*{Response to perpendicular magnetic field}
Figure \ref{PerpField} (b) shows the tunneling conductance $G$ measured as a function of the bias voltage $V$ across Junction 2  with a perpendicular magnetic field $H_\perp$ ranging from 0 to 60 mT. The main feature of the tunneling spectra, strong coherence peaks at $\Delta = \pm 1.25$ meV, is typical for NbSe$_2$~\cite{Noat2015}. We take a closer look at the structure of these peaks by taking the second derivative of the current $d^2I/dV^2$ (Figure \ref{PerpField} (c)). At zero field and positive bias, we see two pairs of peaks, which respond very differently to $H_{\perp}$: while the outer peak only weakly depends on $H_\perp$, the inner peak nearly vanishes at $H_\perp = 60$ mT. Extracting the peak heights vs. vortex separation $r \approx \sqrt{\phi_0/(\pi H_\perp)}$ (with $\phi_0 = h/2e$ the flux quantum) we see that the inner peak disappears at $r \approx 100$ nm.

These results can be understood as signatures of two-band superconductivity, particularly in light of recent theoretical work as well as STM measurements on the paradigmatic two-band superconductor MgB$_2$, which has similar band structure and Fermi surfaces to NbSe$_2$. The zero-field DOS of MgB$_2$ measured by STM exhibits two clearly distinguishable peaks~\cite{Eskildsen:2003kw}. As in our data, a perpendicular magnetic field rapidly smears out the lower-energy peak, whereas the outer peak remains. (These spectra were taken at the superconducting regions, away from vortex cores.) This can be explained by invoking different coherence lengths $\xi_i$, and thus different vortex core sizes, for the two bands ($i = S, L$)~\cite{Nakai:2002fg,Ichioka:2004hf} from which superconductivity arises. The band with the smaller, (partly) induced superconducting gap (denoted $S$) has a larger coherence length and thus a larger vortex core. The $S$ band thus reaches full overlap between vortices well below the critical field of the whole system; at this field, spectral features from the $S$ band are obscured~\cite{Ichioka:2004hf}. Numerical estimates of $\xi_S = \hbar V_\mathrm{F}^S/(\pi \Delta_S)$, for MgB$_2$ agree well with the vortex size of the $S$ band as measured by STM~\cite{Eskildsen:2002ih}.

In the case of NbSe$_2$, data from angle-resolved photoemission spectroscopy (ARPES) give $V_\mathrm{F}^S/V_\mathrm{F}^L \sim 4-5$~\cite{Kiss:2007ky}, consistent with the estimate from our previous study~\cite{Dvir:2018fj}. From the peak locations in Figure \ref{PerpField} (c), we estimate $\Delta_S/\Delta_L \sim 0.6eV/1.2eV \sim 0.5$. As $\xi_i = \hbar V_\mathrm{F}^i/(\pi \Delta_i)$, $\xi_S$ should be $ \approx 160$ nm, i.e. an order of magnitude larger than $\xi_L$, which is known to be $ \approx 20$ nm. This agrees well with the $S$ vortex core size estimated from our data, and with the fact that superconductivity in $S$ is strongly suppressed at a field about two orders of magnitude lower than the upper critical field of NbSe$_2$, which is about $5$T. Our results thus point strongly to superconductivity in NbSe$_2$ arising from two distinct bands with different dynamical properties.

Figure \ref{PerpField} (b) also shows the evolution of sub-gap features in a perpendicular magnetic field. As can be seen here, at zero field, the spectrum shows a hard gap: the low-$V$ conductance is strongly suppressed with respect to that of the normal state ($G_N$). Here, $G$ at $V=0$ (hereafter $G_0$)  $< 0.01 G_N$, making the junctions sensitive to spectral signatures of sub-gap states. Upon application of $H_{\perp}$, $G_0$ rises, while the low-energy spectrum evolves from a ``U''-shape into a ``V" shape (inset). As explicated below, both of these features are associated with CdGM states and point to immediate vortex penetration into the device upon application of $H_{\perp}$. 

\begin{figure}
\includegraphics[width = 0.9\textwidth]{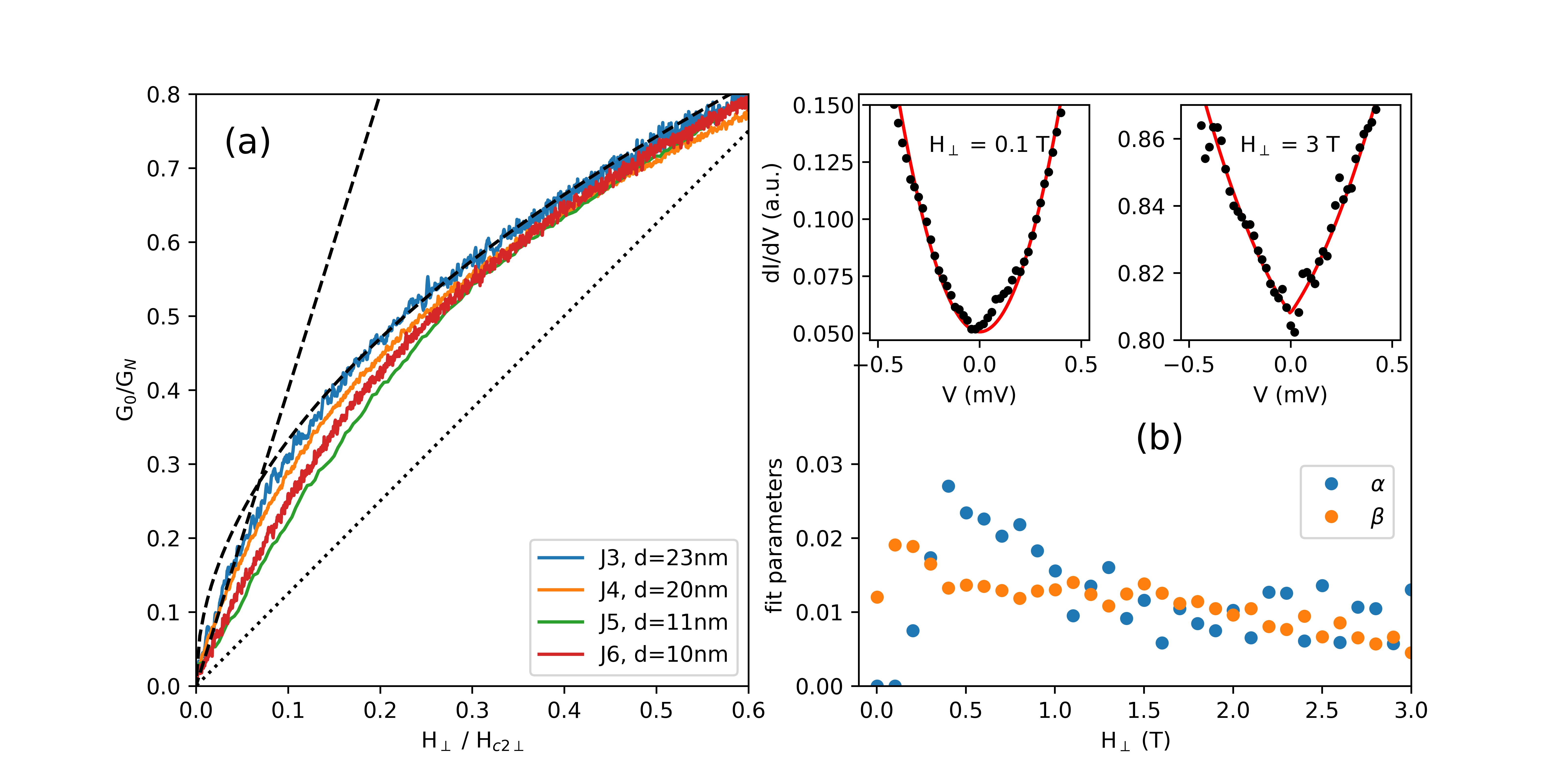}
\caption{Sub-gap spectra at perpendicular magnetic field (a.) $G_0/G_N$ of Junctions 1,3,4,6,7 as a function of $H_\perp$ normalized to the critical field $H_{C2} (T=0)$. Dashed lines shows are a linear curve $G_0/G_n = 4 H_\perp/H_{c2}$, that agrees with the low field dependence and a square root curve $G_0/G_n =  \sqrt{H_\perp/H_{c2}}$ that agrees with the high field data. Dotted line shows are a linear curve $G_0/G_n = 1.25 H_\perp/H_{c2}$ which is expected for a single band isotropic $s$-wave superconductor. (b.) The square ($\beta$) and linear ($\alpha$) fit coefficients extracted from fitting the sub gap spectra of Junction 2 to Equation \ref{sub gap eq.} vs. $H_\perp$. Insets: the sub gap spectra and their respective fits at $H_\perp = 0.1$~T (left) and $H_\perp = 3$~T (right).
}
\label{PerpField_analysis}
\end{figure}

$G_0$ is proportional to the zero-energy DOS, $N_0$. Specific heat measurements in the mixed state~\cite{Hanaguri:2003ce,Nohara:1999bn} have shown that $N_0$ is linearly dependent on $H_\perp$ at low $H_\perp$ and become sub-linear at higher $H_\perp$~\cite{Hanaguri:2003ce}. Our data (Figure~\ref{PerpField_analysis}(a)), measured on five different junctions, follows a similar trend and agrees with early tunneling studies on diffusive superconductors~\cite{Tsuda1969,Guyon:1967do}. We find $G_0 / G_N \sim 4  H_\perp / H_{C2}$  at low $H_\perp$ (<250 mT), higher than expected for an isotropic single-band superconductor \cite{Nakai2004b}.  There are two possible, and non-mutually exclusive explanations for this, and also for the observed highly non-linear $G_0(H_\perp)$ at higher $H_\perp$: gap anisotropy~\cite{Nakai2004b} or two-band superconductivity \cite{Ichioka:2004hf,Nakai:2002fg}. We note that $G_0(H_\perp)$ from Junctions 5 and 6, in which the NbSe$_2$ is thin (sample thickness $d =$ 11-10 nm), exhibit $G_0(H_\perp)$ traces closer to linearity compared to the other, thicker, junctions. This can be interpreted as some level of gap anisotropy being averaged out by a shorter scattering length and leading to a more linear $G_0(H_\perp)$~\cite{Hanaguri:2003ce}. Nevertheless, in all junctions, the $G_0(H_\perp)$ is much further from linearity than what would be expected for a single isotropic band. It is thus likely that both two-band superconductivity and gap anisotropy contribute to the dependence of $G_0$ on $(H_\perp)$ in NbSe$_2$.

Apart from $G_0(H_\perp)$, vortex penetration into the junction area has signatures at finite bias. The area-averaged spectrum of a superconductor in the presence of vortices should assume a linear, V-shaped form, a result found valid for node-less and nodal superconducting order parameters alike~\cite{Nakai2006a}.  This prediction was confirmed experimentally by averaging STM data over a unit cell of the Abrikosov lattice, where the sub-gap spectrum follows an even parabola,
\begin{equation}
G(V) = G_0 (H) + \alpha (H) |V| + \beta (H) V^2. 
\label{sub gap eq.}
\end{equation}
The linear parameter $\alpha (H)$ comes from the spectral weight of area-integrated annular CdGM states, whereas the quadratic term $\beta (H)$ is due to several mechanisms, including thermal smearing, disorder, and Meissner currents~\cite{Nakai2006a}. 

Figure \ref{PerpField_analysis}(b) shows the field dependent coefficients $\alpha(H_\perp)$ and $\beta(H_\perp)$ for Junction 3. We can see that the sub-gap spectrum gradually develops into a V shape, evident in the increase of the linear parameter $\alpha (H_\perp)$ up to $H_\perp\approx0.5$~T. At higher fields, the spectrum remains V-shaped with a decreasing slope, resulting from the increase in the zero bias conductance. This response to field is in qualitative agreement with theory~\cite{Nakai2006a}. The two insets, which compare the sub-gap spectrum at low and high fields, clearly demonstrate this transition from a U shape to a V shape spectrum, indicating the entry of vortices.

\begin{figure}
\includegraphics[width = 0.9\textwidth]{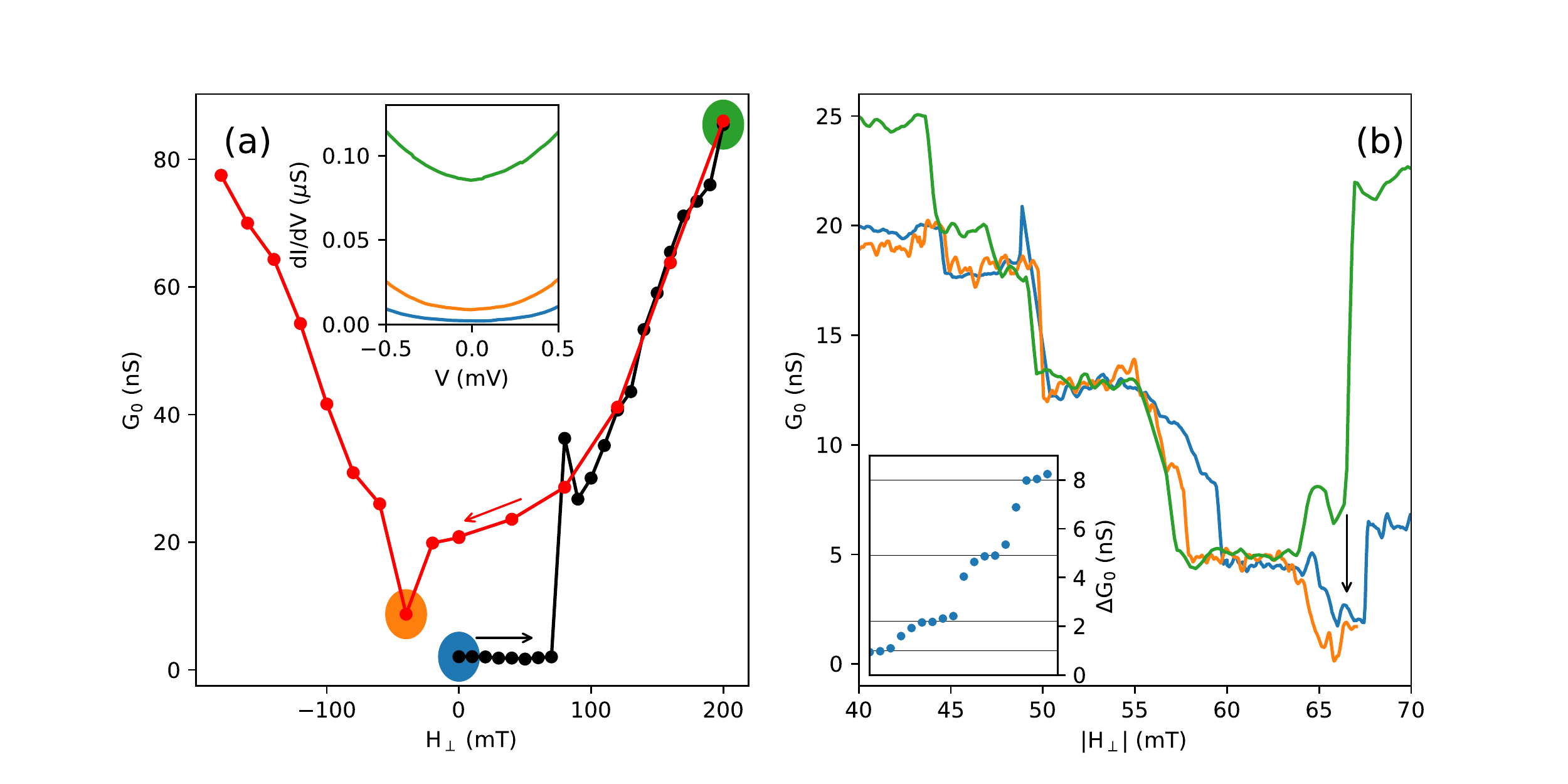}
\caption{Observation of discrete out of plane vortices in planar tunnel junctions. (a.) $G_0$ vs. $H_\perp$ of Junction 8 ($d$ = 3 nm), with $H_\perp$ swept in positive and negative directions (sweep direction marked by an arrow). Inset: Sub-gap spectra, taken at the the marked points in the panel. Curve color corresponds to the color of the marked points. (b.) Multiple scans of $G_0$ as a function of  the absolute value of $H_\perp$, showing conductance plateaus at certain fields. The orange curve is shifted vertically by 4.5~nS and the green curve is shifted horizontally by 9~mT to allow comparison between the conductance plateaus. Inset: The magnitude of differences between the conductance plateaus ($\Delta G_0$), sorted by magnitude.  These measurements were conducted at $^3$He cryostat with a base temperature of 300 mK.
}
\label{DiscreteVrotices}
\end{figure}

\subsection*{Tunneling signature of discrete vortices}

Discrete vortex behavior is detectable in some of our tunnel junctions at low magnetic fields. In Figure \ref{DiscreteVrotices} we plot $G_0$ vs. $H_\perp$ of Junction 8, where $d \approx 3$ nm. In this junction, $G_0$ increases at $H_\perp > $ 60 mT when sweeping the field at a positive direction (panel (a), black). When $H_\perp$ is swept in the opposite direction (red), $G_0$ exhibits hysteresis below 60 mT and attains a minimal value at a negative field (orange dot). The sub-gap spectrum at this field (inset, orange line) differs from that at zero field (blue line), indicating the presence of trapped flux. This can be understood as the result of a surface barrier which is formed on the edge of superconductors \cite{Bean:1964wr}. 
Panel (b) shows similar data taken at higher field resolution, (plotted vs. the absolute value of the field to compare different field sweep directions). Here, $G_0$ exhibits sudden jumps between a series of plateaus. The differences between these plateaus, shown in the inset, are scattered about discrete values ($\Delta G_0 = $ 1, 2.2, 4.9 and 8 nS). We interpret these steps as the entry and exit of a discrete number of vortices into the area probed by the junction. At the field marked by an arrow, the tunneling signal is likely due to a single vortex trapped within the entire junction area. 

\begin{figure}
\includegraphics[width = 0.9\textwidth]{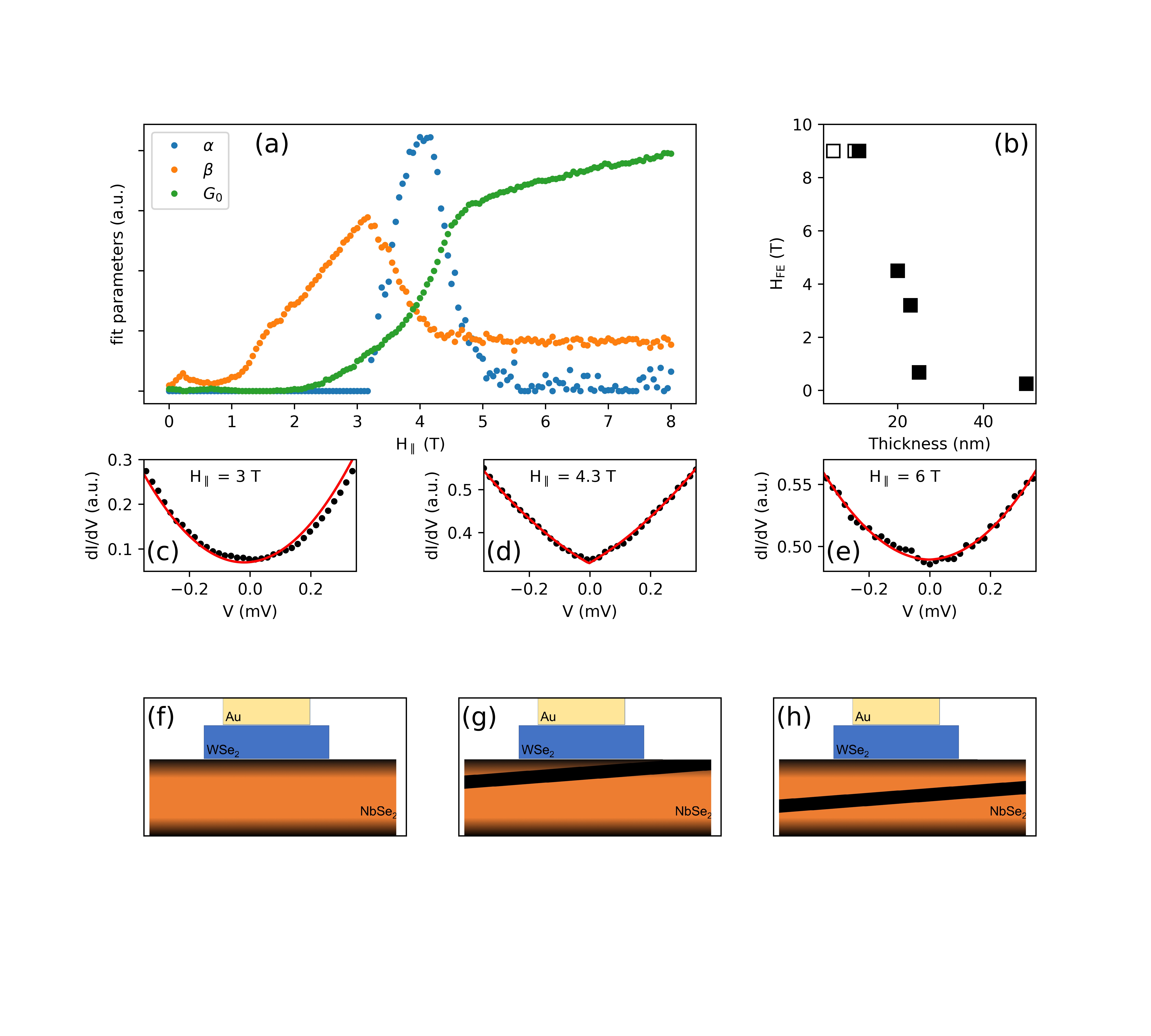}
\caption{Analysis of the spectral response of NbSe$_2$ to in-plane magnetic field.  (a.) The square ($\beta$), linear ($\alpha$) and zero bias ($G_0$) coefficients  resulting from fitting the sub gap spectrum of junction 4 ($d$ = 20 nm) to a parabola as a function of 
$H_\parallel$.  (b.) The penetration field ($H_\mathrm{FE}$) for samples of different thickness. Open squares shows that $H_\mathrm{FE}$ was not reached at 9~T, the maximal field applied.  (c. - e.) The sub gap spectrum and the fit to a parabola of junction 3 at 3~T ($H<H_\mathrm{FE}$), 4.3~T ($H\approx H_\mathrm{FE}$) and 6~T ($H>H_\mathrm{FE}$). (f. - h.) Schematics of the penetration of vortices in in-plane field: at low fields (e) the order parameter (orange) is constant throughout the sample, except at the sample edge; at higher fields, approaching the penetration field $H_\mathrm{FE}$ (g.), vortices nucleate and are bound to surface defects; when the field is high enough ($H>H_\mathrm{FE}$) (h.), vortices enter the bulk of the sample. Note that the field orientation is not perfectly planar. 
}
\label{ParField_analysis}
\end{figure}

\subsection*{Penetration of vortices in parallel magnetic field}

We now turn to the evolution of the sub-gap spectra at parallel magnetic field $H_\parallel$, analysing the spectra using the decomposition used above (Eq. \ref{sub gap eq.}), yielding $G_0(H_\parallel)$, $\alpha(H_\parallel)$ and $\beta(H_\parallel)$. 

Figure \ref{ParField_analysis}a shows the field dependence of the fitting parameters extracted from measurements on Junction 4 ($d = 20$~nm). Upon the application of $H_\parallel$, the first spectral response is an increase of $\beta(H_\parallel)$ (panel c) due to the presence of Meissner currents (panel f). At $H_\parallel\approx$~4T there is a sharp onset of $\alpha(H_\parallel)$ and G$_0(H_\parallel)$ (panel d). This is due to the nucleation of vortices near the surface. Due to slight misalignment, some vortex cores may cross the surface (panel g); these will also contribute $\alpha(H_\parallel)$ and G$_0(H_\parallel)$~\cite{Galvis:2017wm}. Finally, as $H_\parallel$ increases further, $\alpha(H_\parallel)$ is suppressed and $\beta(H_\parallel)$, while reduced, dominates the sub-gap spectrum (panel e). This corresponds to the removal of the energy barrier for vortex penetration into the bulk of the sample \cite{Sutton:1966gh}. This is also accompanied by a kink in $G_0$. This ``first entry'' field $H_{FE}$~\cite{Guyon:1967do} depends on the sample thickness, and should increase with decreasing $d$~\cite{DeGennes:1965bc}. This is the case for our junctions (panel b). In devices of ultra-thin NbSe$_2$, $d\le 5$~nm $\approx \xi_\perp$,  vortices do not penetrate the samples in the entire range of fields that was measured (up-to 9~T). The spectral response of ultra-thin  NbSe$_2$ to parallel field was discussed elsewhere \cite{Dvir:2018fj}. 

\begin{figure}
\includegraphics[width = 0.5\textwidth]{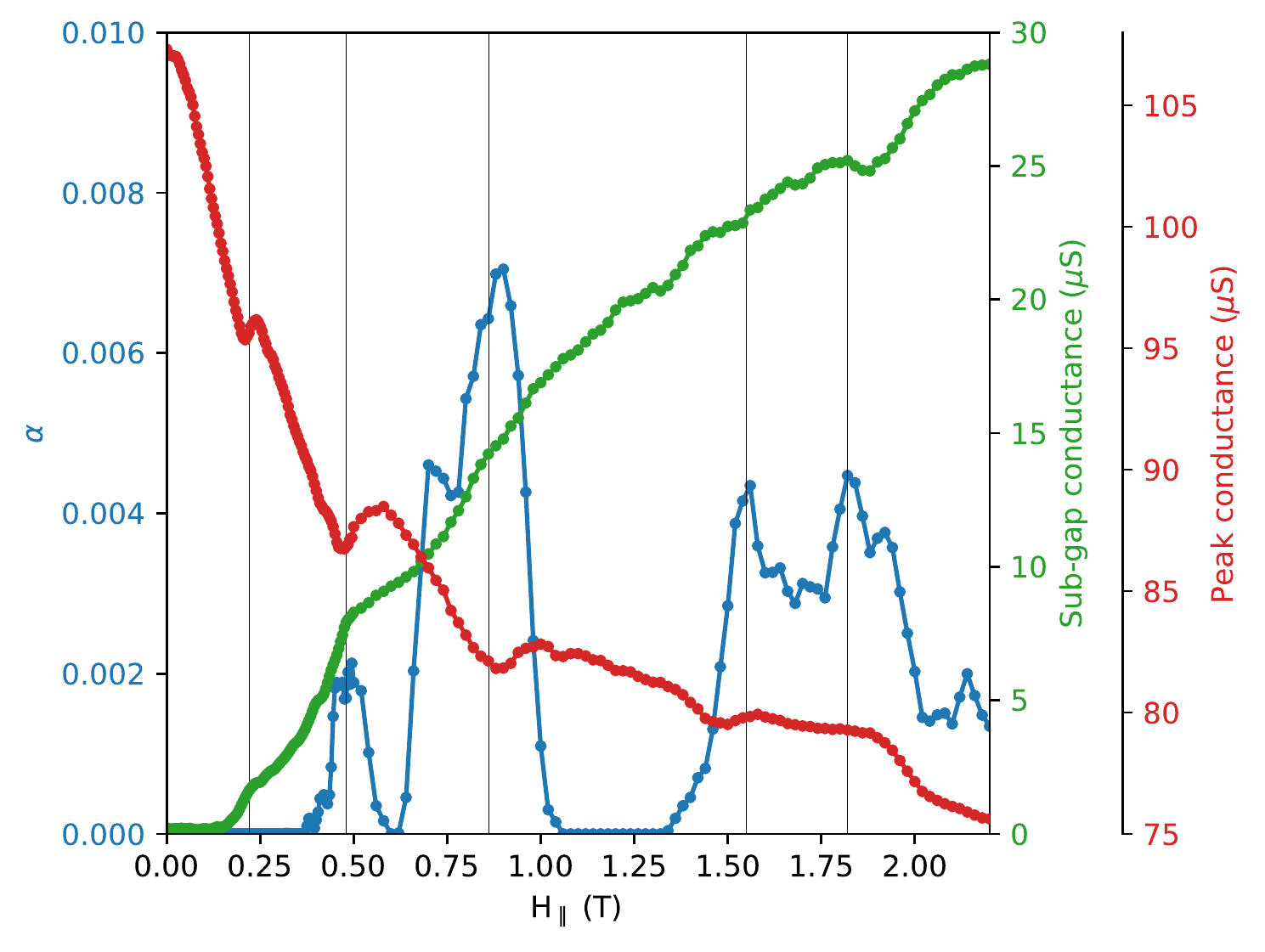}
\caption{Response of a thicker sample ($d \approx 50$ nm) to the application of an in-plane magnetic field. The linear ($\alpha$) term of the subgap spectrum as a function of field (blue) is compared to the subgap conductance ($G_0$) (green) and the maximum conductance of the quasi-particle peaks (red). Black lines serve as a guide to the eye to trace the position of the local minima of peak conductance or maxima of the linear parameter.  }
\label{MiltiEntry}
\end{figure}

\subsection*{Multiple entries of vortices}
As seen in Figure~\ref{MiltiEntry}, in the thickest sample measured (Junction 1, $d =$ 50 nm), $\alpha$ varies non-monotonically between zero and finite values as $H_\parallel$ increases. We interpret this as due to a repeated entry of vortices, which nucleate near the surface and are subsequently pushed to the bulk of the flake. According to theory, in a thin film in a parallel magnetic field above $H_{FE}$, vortices initially form a 1D row within the sample, preventing further flux from entering. At higher fields, more of these rows are formed, with the vortex configuration changing discretely~\cite{Carter:1969gs,Luzhbin:2011ed}. Such discrete changes were observed in measurements of the subgap conductance of thin films of PbIn~\cite{Sutton:1966gh}. Supporting this interpretation, we see that the peaks in $\alpha$ are correlated with kinks in $G_0$, and dips in the height of the quasi-particle peaks. The latter are likely since each entry of vortices suppresses the circulating Meissner currents at the surface.

\subsection*{Conclusion}

Our results demonstrate the utility of tunnel junctions in probing the vortex state in NbSe$_2$. The high spectral resolution allows the differentiation of the two components of the quasiparticle peak. We find the lower energy peak decays at a very small magnetic field far below $H_{C2}$, consistent with a longer coherence length and hence a larger vortex core. At low bias voltages, the hard gap property allows for the observation of the spectral weight of vortex-bound states. We thus find the tunneling signal to be sensitive to the entry of discrete vortices in perpendicular magnetic fields, and to the complex rearrangement of vortices in thin samples in parallel magnetic fields. We suggest that such devices have the potential to probe vortex-bound state in a variety of systems - including, potentially, Majorana states in proximitized topological insulators ~\cite{Fu2008,Xu2015,Sun2016}, and topological superconductors such as FeTe$_{0.55}$Se$_{0.45}$ \cite{Wang:2018ho}.   

\section*{Acknowledgements}
This work was funded by a Maimonïdes-Israel grant from the Israeli-French High Council for Scientific \& Technological Research, an ANR JCJC grant (SPINOES) from the French Agence Nationale de Recherche, and a European Research Council Starting Grant (No. 637298, TUNNEL). T.D. is grateful to the Azrieli Foundation for an Azrieli Fellowship. 

\section*{Author contributions}
T.D. fabricated the devices. C.Q.H.L., T.D. and M.A. performed the measurements. All authors contributed to data analysis and the writing of the manuscript.

\section*{Competing financial interests}
The authors declare no competing financial interests.

\end{document}